\documentclass[aps,prb,reprint,groupedaddress]{revtex4-2}

\usepackage{graphicx}
\usepackage{CJKutf8}
\usepackage{epstopdf}
\usepackage{subfigure}
\usepackage{pifont}
\usepackage{bm}
\usepackage{xcolor}
\usepackage[colorlinks=true,linkcolor=blue,anchorcolor=blue,citecolor=blue,urlcolor=blue]{hyperref}

\begin{document}
\begin{CJK}{UTF8}{gbsn}

\title{Interlayer Pairing Induced Partially Gapped Fermi Surface in Trilayer La$_4$Ni$_3$O$_{10}$ Superconductors}

\author{Junkang Huang}
\author{Tao Zhou}
\email{Corresponding author: tzhou@scnu.edu.cn}

\affiliation{Guangdong Provincial Key Laboratory of Quantum Engineering and Quantum Materials, School of Physics, Guangdong-Hong Kong Joint Laboratory of Quantum Matter, and Frontier Research Institute for Physics, South China Normal University, Guangzhou 510006, China
}

\date{\today}

\begin{abstract}
We explore the superconducting pairing mechanisms in the trilayer $\mathrm{La}_4\mathrm{Ni}_3\mathrm{O}_{10}$ material through self-consistent mean-field calculations. Our findings demonstrate that intralayer pairings are substantially weaker compared to interlayer ones. Remarkably, in the state characterized by interlayer pairing, we detect the presence of partially gapped Fermi surfaces, a fascinating occurrence attributable to the disparity between the inner and outer conducting layers of $\mathrm{La}_4\mathrm{Ni}_3\mathrm{O}_{10}$. Moreover, this study provides valuable insights into the lower superconducting transition temperatures observed in $\mathrm{La}_4\mathrm{Ni}_3\mathrm{O}_{10}$ compounds. This contributes to a deeper understanding of its distinct superconducting attributes.

\end{abstract}
\maketitle

The discovery of high-temperature bilayer nickelate superconductors, $\mathrm{La}_3\mathrm{Ni}_2\mathrm{O}_7$, exhibiting a superconducting transition temperature ($T_c$) of 80 K under pressure, has garnered significant attention \cite{s41586-023-06408-7}. In the domain of cuprate-based high-$T_c$ superconductors, there exists a wide variety of compounds, with trilayer structures achieving the highest superconducting transition temperatures~\cite{PhysRevLett.64.2827,SUN1994122}. This paves the way for the investigation of novel layered nickelate materials that can support high-$T_c$ superconductivity. The potential for superconductivity within trilayer nickelate $\mathrm{La}_4\mathrm{Ni}_3\mathrm{O}_{10}$ was first explored by Sakakibara {\it et al.}, who postulated superconductivity based on density functional theory and fluctuation exchange approximation calculations~\cite{arXiv2309.09462}.

Subsequently, the presence of superconductivity in $\mathrm{La}_4\mathrm{Ni}_3\mathrm{O}_{10}$ was confirmed by various research groups, demonstrating $T_c$ values within the range of $20-30$ K~\cite{0256-307x/41/1/017401,arXiv2311.07353,arXiv2311.07423,arXiv2311.16763}. Notably, $\mathrm{La}_4\mathrm{Ni}_3\mathrm{O}_{10}$ transitions from a monoclinic $P2_1/a$ space group to a tetragonal $I4/mmm$ space group under pressure \cite{arXiv2311.16763}, a transformation similarly observed in $\mathrm{La}_3\mathrm{Ni}_2\mathrm{O}_7$ and considered crucial for the emergence of high-temperature superconductivity \cite{s41586-023-06408-7}. The electronic structures of these compounds exhibit notable similarities, with the Ni-$d_{z^2}$ and Ni-$d_{x^2-y^2}$ orbitals primarily defining the Fermi level, leading to qualitatively similar Fermi surfaces \cite{PhysRevLett.131.126001,PhysRevLett.131.206501,leonov2024electronic,arXiv2402.07196,arXiv2402.02351,arXiv2402.02581,arXiv2402.05085,arXiv2402.05285,arXiv2402.05447,arXiv2402.07196,arXiv2402.07902}. However, unlike the cuprate family, the $T_c$ of the trilayer nickelate $\mathrm{La}_4\mathrm{Ni}_3\mathrm{O}_{10}$ is significantly lower than that of its bilayer counterpart, $\mathrm{La}_3\mathrm{Ni}_2\mathrm{O}_7$.

Comprehensive theoretical and experimental studies have been conducted on $\mathrm{La}_3\mathrm{Ni}_2\mathrm{O}_7$ and $\mathrm{La}_4\mathrm{Ni}_3\mathrm{O}_{10}$. A critical question concerns the nature of interactions responsible for superconductivity. It has been proposed that, unlike conventional layered superconductors where pairing predominantly occurs within the layers, in $\mathrm{La}_3\mathrm{Ni}_2\mathrm{O}_7$, strong interlayer hopping of $d_{z^2}$ orbitals suggests that interlayer interactions might significantly influence pairing. Consequently, the dynamics between intra- and interlayer pairing has become a focal point~\cite{arXiv2306.07275,wu2024,Shen_2023,PhysRevB.108.L140504,lu2023superconductivity,lange2024pairing,schlomer2023superconductivity,PhysRevB.108.L201108,arXiv2310.02915,PhysRevB.108.174511,arXiv2311.05491,PhysRevLett.132.146002,arXiv2402.07449,PhysRevB.108.165141,PhysRevLett.131.236002,PhysRevLett.132.036502,fan2023superconductivity,arXiv2309.05726,arXiv2309.15095,PhysRevLett.132.106002,PhysRevB.108.174501,arXiv2401.15097,arXiv2308.16564,PhysRevB.108.214522,arXiv2310.17465,arXiv2308.09698}, with some studies suggesting that interlayer pairing plays a dominant role in superconductivity~\cite{PhysRevB.108.L201108,arXiv2310.02915,PhysRevB.108.174511,arXiv2311.05491,PhysRevLett.132.146002,arXiv2402.07449}, enhanced by significant Hund's coupling and fostering substantial interlayer interactions~\cite{PhysRevB.109.L081105,arXiv2311.12769,arXiv2308.09698,arXiv2310.02915,PhysRevB.108.174511}.
Experimental results from neutron scattering and resonant inelastic X-ray scattering further imply that magnetic superexchange interactions between layers considerably exceed those within layers.~\cite{xie2024neutron,chen2024electronic}.
Recent interest in the material $\mathrm{La}_4\mathrm{Ni}_3\mathrm{O}_{10}$ has been largely focused on its pairing mechanism and symmetry~\cite{leonov2024electronic,arXiv2402.07196,arXiv2402.02351,arXiv2402.02581,arXiv2402.05085,arXiv2402.05285,arXiv2402.05447,arXiv2402.07196,arXiv2402.07902}. Despite this attention, the intricate relationship and competition between interlayer and intralayer interactions are yet to be fully explored. Investigating these aspects could clarify the nuances of its superconducting properties and potentially highlight paths to enhance its performance.

The notable decrease in $T_c$ for the trilayer compound has generated theoretical scrutiny. The proposed explanations include diminished electronic correlation in $\mathrm{La}_4\mathrm{Ni}_3\mathrm{O}_{10}$ due to increased hole doping \cite{arXiv2402.02581}, a reduced pairing eigenvalue linked to weak antiferromagnetic exchange between the top and bottom layers \cite{arXiv2402.05447,arXiv2402.07902}, and interlayer spin antiferromagnetic exchange catalyzed by interlayer hopping \cite{arXiv2402.06450}. Notably, in multi-layer cuprate-based superconducting materials, $T_c$ significantly increases with the layer count up to three and then gradually declines~\cite{PhysRevLett.64.2827,SUN1994122}. This distinct difference between nickelates and cuprates offers an intriguing insight and could be crucial in understanding the nature of superconductivity in layered nickelate materials.

In this paper, we explore the superconductivity of $\mathrm{La}_4 \mathrm{Ni}_3 \mathrm{O}_{10}$ through a two-orbital trilayer tight-binding model and a self-consistent approach. Our numerical explorations reveal that interlayer pairings predominantly facilitate superconductivity, whereas intralayer pairing magnitudes are negligibly small. In the superconducting state, the normal state Fermi surfaces are incompletely gapped, leaving some segments/points ungapped. This manifestation of partial Fermi surfaces in the superconducting state is ascribed to the energy band discrepancies between inner and outer layer quasiparticles, thereby offering a plausible explanation for the diminished T$_c$ in the $\mathrm{La}_4 \mathrm{Ni}3 \mathrm{O}{10}$ material.


Our investigation begins with a two-orbital model situated on a trilayer square lattice, encapsulating both the tight-binding and interaction terms, formulated as,
\begin{eqnarray}
	H = -\sum_{l,l'}\sum_{ij\tau\tau'\sigma} t_{ij\tau\tau'}^{l,l'}c_{i\tau\sigma}^{l\dagger}c_{j\tau'\sigma}^{l'\dagger} +H_I.
\end{eqnarray}
Here $l$ and $l'$ represent the layer indices, while $\tau$ and $\tau'$ denote the orbital indices, and $\sigma$ refers to the spin index. The tight-binding parameters $t_{ij\tau\tau'}^{l,l'}$
 are obtained from Ref. \cite{arXiv2402.07196}. $H_I$ signifies the superconducting pairing term, detailed as $H_I=\sum_{l,l^\prime}\sum_{ij\tau}V_{ij\tau}^{l,l'}( c_{i\tau\uparrow}^{l\dagger} c_{j\tau\downarrow}^{l^\prime\dagger} c_{i\tau\uparrow}^{l} c_{j\tau\downarrow}^{l^\prime}+c_{i\tau\downarrow}^{l\dagger} c_{j\tau\uparrow}^{l^\prime\dagger} c_{i\tau\downarrow}^{l} c_{j\tau\uparrow}^{l^\prime})$. The mean-field order parameters for intra- and interlayer pairings are defined as  $\Delta_{ij\tau}^{l,l} = \frac{V_{ij\tau}^{l,l}}{2} \left \langle c_{i\tau\uparrow}^{l} c_{j\tau\downarrow}^{l} - c_{i\tau\downarrow}^{l} c_{j\tau\uparrow}^{l} \right \rangle$, $\Delta_{ii\tau}^{l,l'} = \frac{V_{ii\tau}^{l,l'}}{2} \left \langle c_{i\tau\uparrow}^{l} c_{i\tau\downarrow}^{l'} - c_{i\tau\downarrow}^{l} c_{i\tau\uparrow}^{l'} \right \rangle$, respectively.
 
Through Fourier transformation, the Hamiltonian can be expressed in momentum space as $H = \sum_{\bf{k}} \Psi^{\dagger}\left(\bf{k}\right) \hat{M}\left(\bf{k}\right) \Psi\left(\bf{k}\right)$, where the vector $\Psi\left(\bf{k}\right) = \left( u_{\bf k}, v_{\bf k} \right)$ is given by
\begin{eqnarray}
	u_{\bf k} &=& \left( c_{\bf{k}1\uparrow}^{1}, c_{\bf{k}2\uparrow}^{1}, c_{\bf{k}1\uparrow}^{2}, c_{\bf{k}2\uparrow}^{2}, c_{\bf{k}1\uparrow}^{3}, c_{\bf{k}2\uparrow}^{3} \right)^T, \nonumber \\
	v_{\bf k} &=& \left( c_{-\bf{k}1\downarrow}^{1\dagger}, c_{-\bf{k}2\downarrow}^{1\dagger}, c_{-\bf{k}1\downarrow}^{2\dagger}, c_{-\bf{k}2\downarrow}^{2\dagger}, c_{-\bf{k}1\downarrow}^{3\dagger}, c_{-\bf{k}2\downarrow}^{3\dagger} \right)^{T},
\end{eqnarray}
and $\hat{M}\left(\bf{k}\right)$ is a $12 \times 12$ matrix.

The superconducting pairing terms include the intralayer channel and interlayer channel. Specifically, for intralayer attractions, we focus on nearest-neighbor interactions with $V_{\tau\parallel}=V_{ij\tau}^{l,l}$,  representing the extended $s$-wave order parameters of the outer (O) (l=1,3) and inner (I) (l=2) layers. These are determined self-consistently as,
\begin{eqnarray}
	\label{EQ:self1}
	\Delta_{\tau}^{O\left(I\right)} = \frac{V_{\tau\parallel}}{4N} \sum_{n\bf{k}} \left( \cos \bf{k}_x + \cos \bf{k}_y \right) u_{\tau n\bf{k}}^{O\left(I\right)*} v_{\tau n\bf{k}}^{O\left(I\right)} \tanh \frac{\beta E_{n\bf{k}}}{2}.
\end{eqnarray}

For the interlayer interaction, characterized by the interlayer interaction parameter $V_{\tau\perp}=V_{ii\tau}^{l,l+1}$, the order parameters $\Delta_{\tau\perp}$ are similarly determined,
\begin{eqnarray}
	\label{EQ:self2}
	\Delta_{\tau\perp} = \frac{V_{\tau\perp}}{2N} \sum_{n\bf{k}} u_{\tau n\bf{k}}^{O*}v_{\tau n\bf{k}}^{I} \tanh \frac{\beta E_{n\bf{k}}}{2},
\end{eqnarray}
where $E_{n\textbf{k}}$ denotes the eigenvalues of $\hat{M}(\textbf{k})$ and $\beta = 10^{-5}$.

The spectral function, a crucial quantity for probing the electronic structure, is computed as,
\begin{eqnarray}
	A\left(\bf{k},\omega\right) = -\frac{\mathrm{Im}}{\pi} \sum_{p=1}^6 \sum_n \frac{\left| u_{p,n\bf{k}} \right|^2}{\omega - E_{n\bf{k}} + i\Gamma} + \frac{\left| v_{p,n\bf{k}} \right|^2}{\omega + E_{n\bf{k}} + i\Gamma},
\end{eqnarray}
 with a damping factor $\Gamma = 0.002$. 

Lastly, the density of states is represented as a function of frequency via integration of the spectral function over all momenta,
\begin{eqnarray}
	\rho\left( \omega \right) = \frac{1}{N} \sum_{\bf k} A \left( \bf k,\omega\right).
\end{eqnarray}

\begin{figure}[bp]
	\centering
	\includegraphics[width = 8cm]{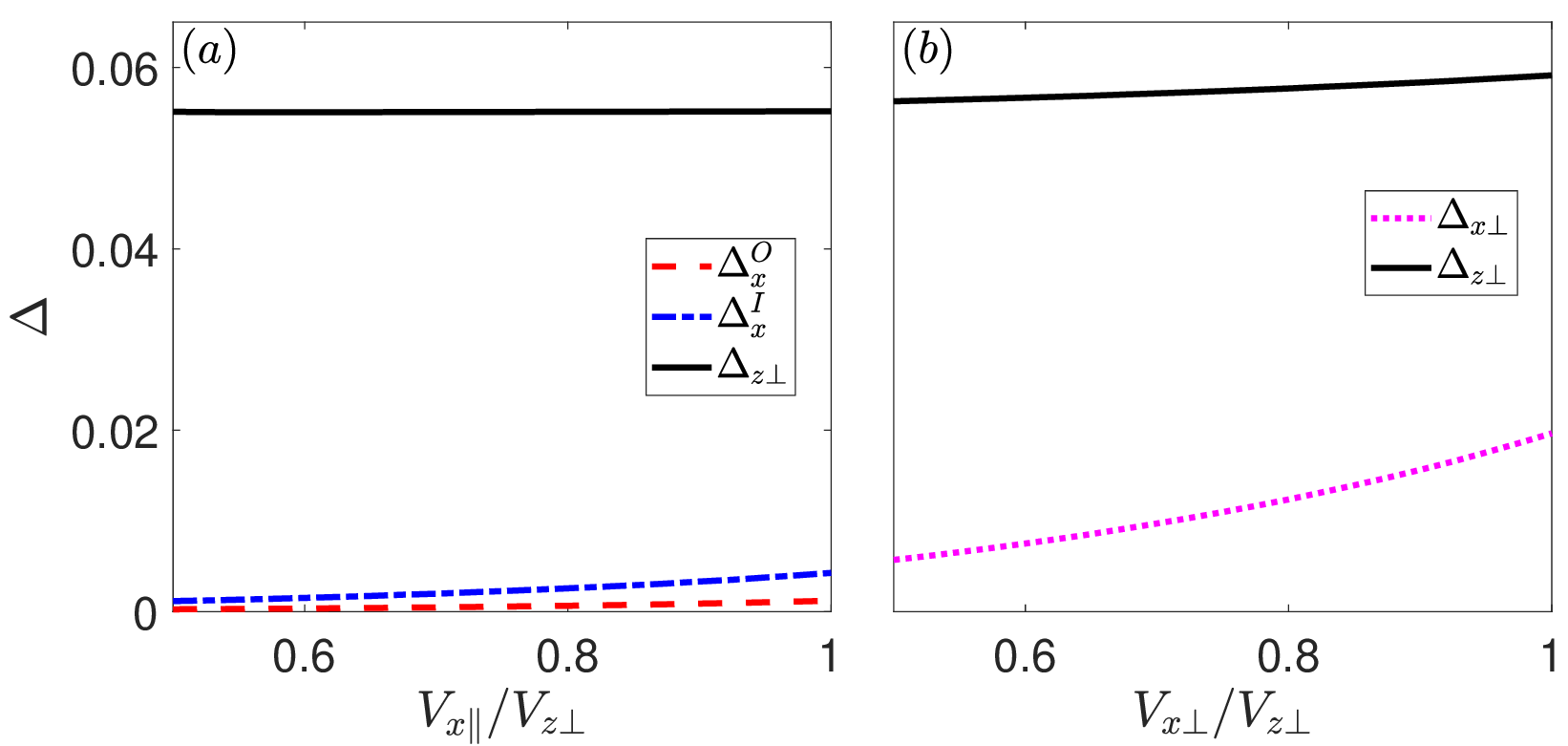}
	\caption{\label{fig:self} Variation of order parameters with pairing potential from self-consistent calculations}
\end{figure}

We begin our discussion by exploring the underlying interactions responsible for superconductivity. At the microscopic level, superconducting pairing is attributed to superexchange interactions $J$ with $J\approx 4t^2/U$. Consequently, the effective pairing strength $V$, is fundamentally related to the square of the hopping constant, denoted as $V \propto t^2$. Here, $t$ represents the hopping constant, as defined in the Hamiltonian of Eq.(1), derived from density functional theory calculations specified by Ref.~\cite{arXiv2402.07196} for pressurized $\mathrm{La}_4 \mathrm{Ni}_3 \mathrm{O}_{10}$. The calculated interlayer hopping constant ($t_\perp$) for the $d_{z^2}$ orbital is 0.738, while the intralayer hopping constants ($t_\parallel$) for the outer and inner layers of the $d_{x^2-y^2}$ orbital are 0.511 and 0.521, respectively. Hopping constants through other channels are significantly lesser in magnitude. Subsequently, we estimate the ratio of the intralayer to interlayer pairing strength as $V_{x\parallel}/V_{z\perp} \approx 0.5$.

We now study the competition of the interlayer pairing and intralayer pairing numerically. Fixing the interlayer pairing strength for the $d_{z^2}$ orbital $V_{z\perp}$ at $0.8$ for illustrative purposes, we utilize a self-consistent method (Eqs. [\ref{EQ:self1}, \ref{EQ:self2}]) to derive the behavior of order parameters relative to the intralayer pairing strength, $V_{x\parallel}$. As we adjust $V_{x\parallel}$ from $0.4$ to $0.8$, Fig. \ref{fig:self}(a) reveals that the order parameter associated with interlayer pairing is largely unaffected by variations in $V_{x\parallel}$, maintaining a consistently high value. Remarkably, at a ratio of $V_{x\parallel}/V_{z\perp} = 0.5$, the magnitudes of the intralayer order parameters for both layers diminish to nearly zero. Further increments in $V_{x\parallel}$ result in a slight increase in the intralayer order parameters; however, these remain significantly lower than the interlayer parameter, even as $V_{x\parallel}$ equals $V_{z\perp}$. This observed trend underscores the predominance of interlayer pairing in determining the superconductivity of $\mathrm{La}_4\mathrm{Ni}_3\mathrm{O}_{10}$, a conclusion strongly supported by our numerical simulations, which were specifically conducted under the presumption of a $V_{x{\parallel}}/V_{z\perp} \approx 0.5$ interlayer pairing ratio.

\begin{figure}
	\centering
	\includegraphics[width = 8cm]{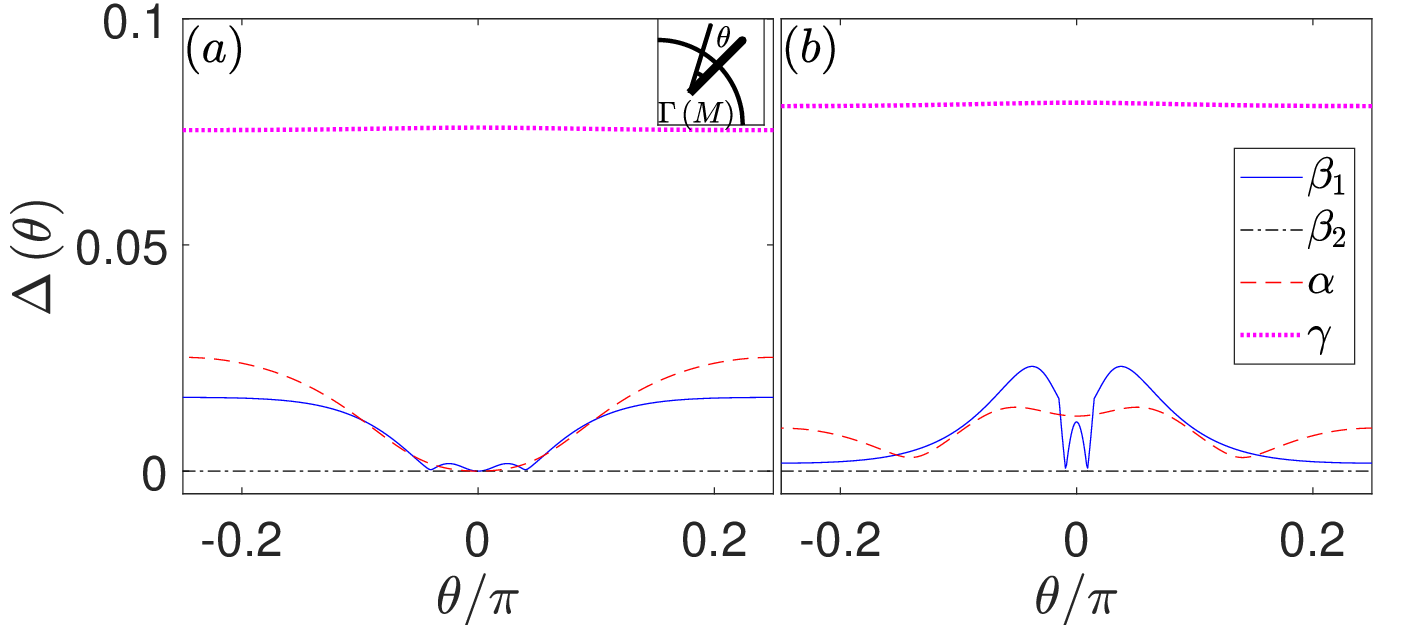}
	\caption{\label{fig:gapFP} Energy gap magnitudes along different Fermi pockets (a) without and (b) with interlayer $d_{x^2-y^2}$ orbital pairing.
}
\end{figure}

The interlayer hopping constant for the $d_{x^2-y^2}$ orbital is notably small~\cite{arXiv2402.07196}, essentially nullifying direct superexchange interactions for this orbital. Nevertheless, due to substantial Hund's coupling, the $d_{z^2}$ orbital can endow the $d_{x^2-y^2}$ orbital with an interlayer pairing potential, thereby enabling interlayer pairing in the latter\cite{arXiv2310.02915,PhysRevB.108.174511,arXiv2311.05491,PhysRevLett.132.146002,arXiv2402.07449}. In this context, with the recognition of an effective interlayer pairing potential in the $d_{x^2-y^2}$ orbital ($V_{x\perp}$), our numerical analysis delineates the behavior of order parameters as a function of $V_{x\perp}$, which ranges from $0.4$ to $0.8$, as shown in Fig. \ref{fig:self}(b). Remarkably, as $V_{x\perp}$ increases, not only is the pairing order parameter for the $d_{x^2-y^2}$ orbital initiated and bolstered, but there is also a slight elevation in the pairing order parameter of the $d_{z^2}$ orbital. This observation suggests that the presence of Hund's coupling could potentially elevate the superconducting transition temperature.

The normal state Fermi surface of $\mathrm{La}_4 \mathrm{Ni}_3 \mathrm{O}_{10}$ comprises four distinct Fermi pockets, namely $\gamma$, $\beta_1$, and $\beta_2$ pockets encircling the $M=(\pi,\pi)$ point of the Brillouin zone, and the $\alpha$ pocket encircling the $\Gamma=(0,0)$ point~\cite{arXiv2402.07196}. Our investigation explores the energy gaps along this normal state Fermi surface under the influence of interlayer pairing. The numerical results for two configurations, $(V_{x\perp},V_{z\perp})=(0,0.8)$ and $(V_{x\perp},V_{z\perp})=(0.8,0.8)$, are depicted in Fig. \ref{fig:gapFP}(a) and Fig. \ref{fig:gapFP}(b), respectively. With the absence of the $d_{x^2-y^2}$ orbital's pairing term, the most significant gap manifests along the $\gamma$ Fermi pocket, attributable to the flat band proximal to the Fermi level within this band. The energy gaps across this pocket are nearly uniform. Remarkably, the $\beta_2$ Fermi surface exhibits no gaps. Conversely, the energy gaps demonstrated by the $\beta_1$ and $\alpha$ Fermi pockets show variability and display anisotropic properties, particularly, the gap magnitudes diminish to nearly zero in proximity to the diagonal direction and increase progressively as one moves away from this axis.

Upon the introduction of the pairing term for the $d_{x^2-y^2}$ orbital, key features broadly maintain qualitative similarity. The energy gaps along the $\gamma$ Fermi pocket persist in being isotropic and exceedingly large, with slight increments in magnitude. The $\beta_2$ Fermi pocket continues exhibiting a gapless state. However, the anisotropic nature of the energy gaps around the $\beta_1$ and $\alpha$ Fermi pockets evolve; with enhanced pairing in the $d_{x^2-y^2}$ orbital, distinctions in anisotropic behavior emerge. Notably, a minor gap appears precisely at the diagonal orientation ($\theta=0$) of the $\beta_1$ Fermi pocket, with two nodal points positioned nearby. As $\theta$ extends, the energy gap escalates to its peak before diminishing with further increase in $\theta$, ultimately approaching zero as it nears the off-diagonal position. Regarding the $\alpha$ Fermi surface, it becomes fully gapped, although the gap significantly reduces at certain intervals, demonstrating a nuanced modification induced by the addition of pairing in the $d_{x^2-y^2}$ orbital.

\begin{figure}
	\centering
	\includegraphics[width = 8.5cm]{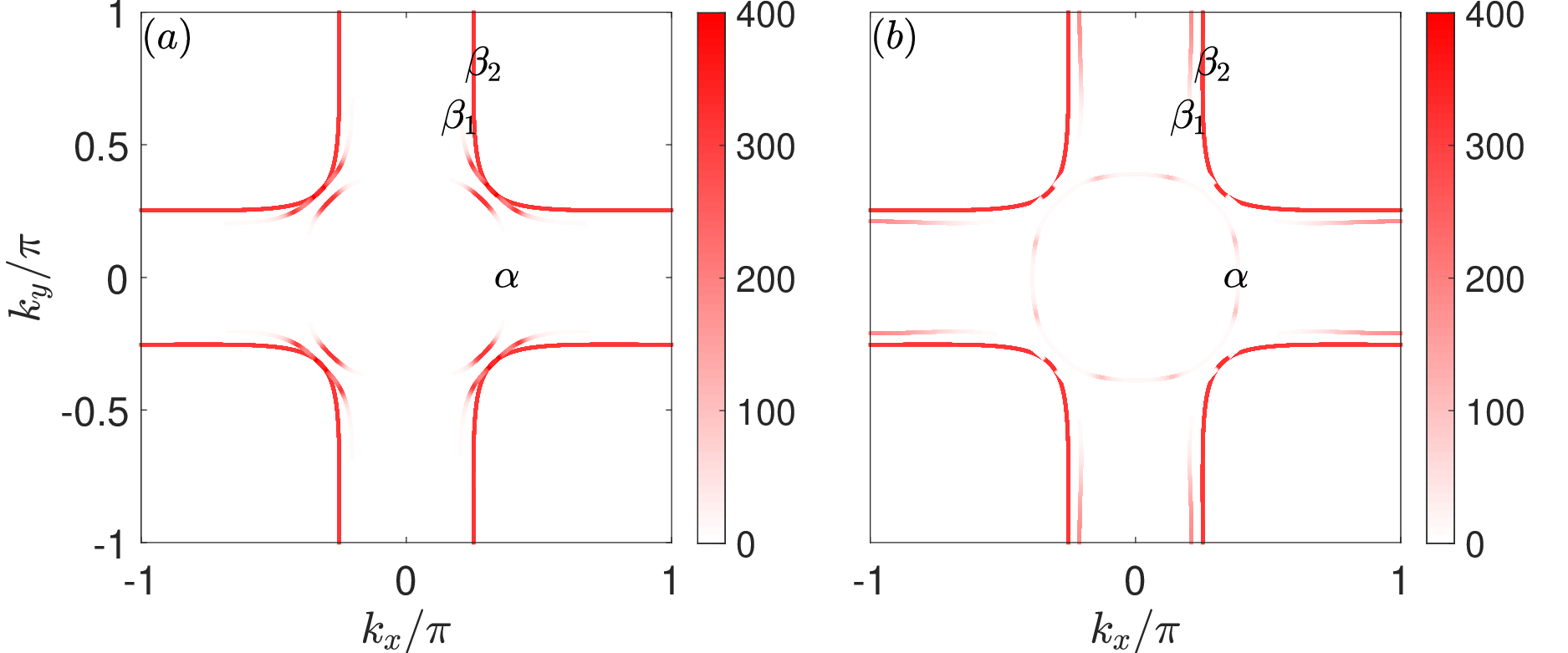}
	\caption{\label{fig:SF} Zero-energy spectral function spectra in the first Brillouin zone (a) without and (b) with interlayer $d_{x^2-y^2}$ orbital pairing.}
\end{figure}

Details regarding the superconducting energy gap can be determined experimentally through techniques such as angle-resolved photoemission spectroscopy (ARPES) and scanning tunneling microscopy (STM). These experimental findings can be theoretically correlated by calculating the spectral function and the density of states. Intensity plots of zero energy spectral function spectra in the superconducting state, for configurations $(V_{x\perp},V_{z\perp})=(0,0.8)$ and $(V_{x\perp},V_{z\perp})=(0.8,0.8)$, are presented in Fig. \ref{fig:gapFP}(a) and \ref{fig:gapFP}(b), respectively.

The zero-energy spectral function is indicative of the distribution of zero-energy quasiparticles, typically found in the gapless regions of the Fermi surface. Around the $\gamma$ Fermi pocket, the spectral function exhibits zero intensity, reflecting the large superconducting gap. Conversely, the entire $\beta_2$ Fermi pocket is characterized by its gapless nature, resulting in pronounced zero-energy spectral functions along this pocket. In the absence of pairing potential in the $d_{x^2-y^2}$ orbital, as shown in Fig. \ref{fig:gapFP}(a), the $\beta_1$ and $\alpha$ Fermi pockets exhibit partial gapping, with segments of the Fermi surface existing along the diagonal direction. With the introduction of pairing potential in the $d_{x^2-y^2}$ orbital, evidenced in Fig. \ref{fig:gapFP}(b), quasiparticles emerge at locations where energy gaps are minimal. Specifically, for the $\beta_1$ pocket, quasiparticles are observed near the Brillouin zone boundary. Although the $\alpha$ Fermi pocket is entirely gapped, minute energy gaps at certain locations result in a non-zero spectral function at these points.

\begin{figure}[tp]
	\centering
	\includegraphics[width = 8cm]{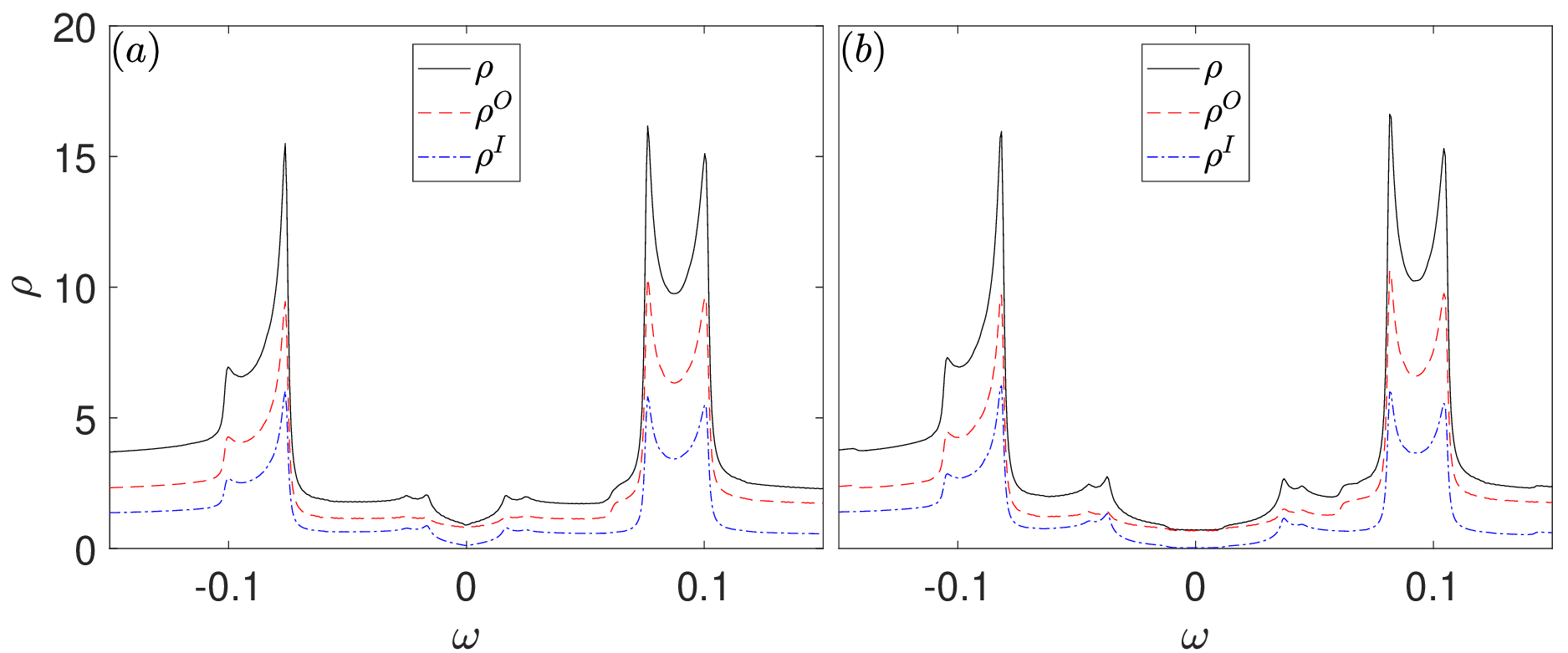}
	\caption{\label{fig:DOS} Density of states (a) in the absence and (b) presence of interlayer $d_{x^2-y^2}$ orbital pairing.}
\end{figure}

We next examine the density of states spectra, also under two different configurations: $(V_{x\perp},V_{z\perp})=(0,0.8)$ and $(V_{x\perp},V_{z\perp})=(0.8,0.8)$. The numerical results are illustrated in Fig. \ref{fig:DOS}. A notable peak in positive energy, approximately at $0.1$, originates from the van Hove singularity in the normal state energy bands. Additional peaks observed are attributable to superconducting coherence peaks, which arise from superconducting pairing.

Distinct two-gap features are evident within the spectra. The larger superconducting coherence peaks, associated with sizeable gaps, are primarily contributions from the $\gamma$ Fermi pocket. Conversely, the $\beta_1$ Fermi pocket is responsible for the smaller energy gaps observed. The spectral weight associated with the $\alpha$ Fermi pocket is comparatively low, resulting in the absence of superconducting coherence peaks attributed to this pocket.

The introduction of pairing in the $d_{x^2-y^2}$ orbital markedly increases the maximum superconducting gap around the $\beta_1$ Fermi pocket, as detailed in Fig. \ref{fig:gapFP}. Consequently, the superconducting coherence peaks, primarily contributed by the $\beta_1$ Fermi pocket, also advance to higher energy levels.

Additionally, our numerical analyses suggest a contrasting behavior between the inner and outer layer density of states near the Fermi level. Specifically, the density of states approaches zero at the Fermi level within the inner layer, yet remains finite at low energies in the outer layer. This implies that zero-energy quasiparticles in the superconducting state are predominantly localized in the outer layer.

\begin{figure}[tp]
	\centering
	\includegraphics[width = 8.5cm]{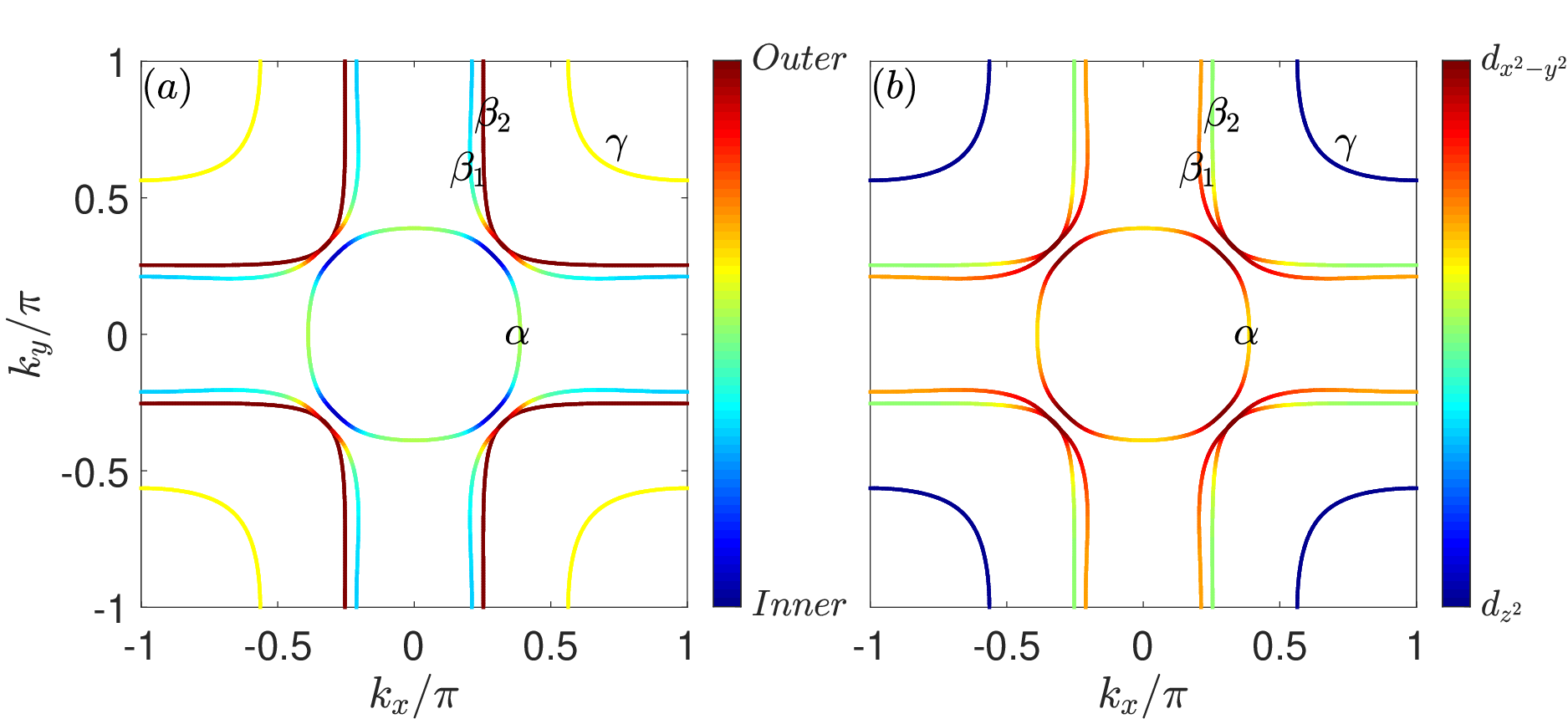}
	\caption{\label{fig:FS2} (a) Normal state Fermi surface of $\mathrm{La}_4\mathrm{Ni}_3\mathrm{O}_{10}$, highlighted by layer contributions. (b) Modified Fermi surface depiction emphasizing orbital weight.}
\end{figure}

The phenomenon of a partially gapless Fermi surface can be coherently explained through the Fermiology inherent to the system. The La$_4$Ni$_3$O$_{10}$ compound is structured with three NiO$_2$ layers within its unit cell, where the outer and inner layers are distinct from each other. Consequently, in the normal state, the distribution of quasiparticles at the Fermi level differs between the outer and inner layers. In the transition to the superconducting state, where pairing occurs between the inner and outer layers, a natural consequence is the inability of some quasiparticles at the Fermi level to pair, resulting in a partially gapless Fermi surface. For a clearer understanding, we have illustrated the normal state Fermi surface in Fig. \ref{fig:FS2}, with color bars in Fig. \ref{fig:FS2}(a) and \ref{fig:FS2}(b) indicating the contributions from the respective layers and orbitals to the Fermi surface. As demonstrated in Fig. \ref{fig:FS2}(a), the $\beta_2$ Fermi pocket is entirely a result of the outer layer contribution, rendering it ungappable in the superconducting state. Conversely, as shown in Fig. \ref{fig:FS2}(b), the $\beta_1$ and $\alpha$ Fermi pockets primarily derive from the $d_{x^2-y^2}$ orbital. Notably, along the diagonal direction, the contribution from the $d_{x^2-y^2}$ orbital reaches its maximum. Hence, in the absence of pairing within the $d_{x^2-y^2}$ orbital, the $\beta_1$ and $\alpha$ Fermi pockets remain gapless along the diagonal direction. The introduction of interlayer pairing within the $d_{x^2-y^2}$ orbital significantly influences the energy gaps of these pockets, with the gap magnitudes strongly correlated to the contribution weight of the $d_{x^2-y^2}$ orbital.

The observed significant reduction in the superconducting transition temperature $T_c$ of $\mathrm{La}_4\mathrm{Ni}_3\mathrm{O}_{10}$ compared to $\mathrm{La}_3\mathrm{Ni}_2\mathrm{O}_{7}$ can be attributed to the distinct layer configurations and their contribution to superconductivity. Specifically, in $\mathrm{La}_4\mathrm{Ni}_3\mathrm{O}_{10}$, the non-equivalence of the inner and outer layers means that only a subset of the quasiparticles at the Fermi level engage in superconductivity within an interlayer pairing framework. In contrast, the bilayer $\mathrm{La}_3\mathrm{Ni}2\mathrm{O}_{7}$ features two equivalent NiO$_2$ layers, enabling all quasiparticles at the Fermi level to participate in superconductivity, thus resulting in a higher $T_c$. This conceptual framework is supported by numerical analyses; using the same pairing interaction as in our study, the order parameter in bilayer samples is roughly twice that observed in our current findings. Given that the magnitude of the superconducting gap is directly proportional to $T_c$ at the mean field level, this approach effectively explains the observed differences in $T_c$ between the bilayer and trilayer nickelate superconductors.

To deepen our understanding, a comparative examination of multilayer cuprate and nickelate superconductors is essential.  In cuprate superconductors, the $d_{x^2-y^2}$ orbital plays a pivotal role in fostering intralayer superconducting pairing. For multilayer samples, it was proposed that $T_c$ can significantly increase as a result of superconducting pairing tunneling between layers~\cite{PhysRevLett.106.167001}. Consequently, for a trilayer sample, a considerable elevation in $T_c$ is observed~\cite{PhysRevLett.64.2827,SUN1994122}. It is important to highlight, however, that for samples with four or more layers, $T_c$ tends to decrease gradually, a phenomenon attributed to inhomogeneous charge distribution\cite{PhysRevLett.64.2827}. In stark contrast, superconductivity in nickelate materials is predominantly reliant on interlayer pairing, with pairing tunneling playing no practical role. Thus, for multilayer nickelates, increasing the number of layers does not correlate with an increase in $T_c$.


In summary, our mean-field level analyses, complemented by self-consistent calculations, lead us to conclude that interlayer pairing mechanisms are responsible for the superconductivity observed in trilayer $\mathrm{La}_4\mathrm{Ni}_3\mathrm{O}_{10}$ material. In the superconducting state, the Fermi surface exhibits partially gapped pockets. The presence of gapless Fermi surface segments can be attributed to the asymmetry between the inner and outer NiO$_2$ layers. This framework also provides a plausible explanation for the observed decrease in the superconducting transition temperature of $\mathrm{La}_4\mathrm{Ni}_3\mathrm{O}_{10}$.


	This work was supported by the NSFC under the Grant No.12074130.
%

\end{CJK}
\end{document}